\documentclass{aastex}
\usepackage{emulateapj5}

\def\agt{\mathrel{\raise.3ex\hbox{$>$}\mkern-14mu\lower0.6ex\hbox{$\sim$}}}
\def\alt{\mathrel{\raise.3ex\hbox{$<$}\mkern-14mu\lower0.6ex\hbox{$\sim$}}}

\newcommand{\beq}{\begin{equation}}
\newcommand{\eeq}{\end{equation}}
\newcommand{\beqn}{\begin{eqnarray}}
\newcommand{\eeqn}{\end{eqnarray}}

\journalinfo{THE ASTROPHYSICAL JOURNAL, 577, 2002 October 1}
\shorttitle{Collapse of Rotating Supermassive Star}
\shortauthors{Shapiro \& Shibata}

\newlength{\minitwocolumn}
\begin{document}

\title{Collapse of a Rotating Supermassive Star to a
Supermassive Black Hole: Analytic Determination of
the Black Hole Mass and Spin}

\author{Stuart L. Shapiro \altaffilmark{1,2}
and Masaru Shibata \altaffilmark{3}}

\affil
{\altaffilmark{1} 
Department of Physics, University of Illinois at Urbana-Champaign,
\break
Urbana, IL 61801-3080}
\affil
{\altaffilmark{2} 
Department of Astronomy and NCSA, University of Illinois at
Urbana-Champaign,
\break
Urbana, IL 61801-3080}

\affil
{\altaffilmark{3} 
Graduate School of Arts and Sciences, 
University of Tokyo,
\break
Komaba, Meguro, Tokyo 153-8902, Japan}

\begin{abstract}
The collapse of a uniformaly rotating, supermassive star (SMS)
to a supermassive black hole (SMBH) has been followed recently
by means of hydrodynamic simulations in full general relativity.
The initial SMS of arbitrary mass $M$ in these simulations 
rotates uniformly at the mass--shedding limit and 
is marginally unstable to radial collapse. 
The final black hole mass and spin
have been determined to be  $M_h/M \approx 0.9$ and
$J_h/M_h^2 \approx 0.75$. The 
remaining mass goes into a disk of mass $M_{\rm disk}/M \approx 0.1$. Here
we show that these black hole and disk parameters can be calculated 
{\it analytically} from the initial stellar density and angular momentum 
distribution. The analytic calculation thereby corroborates and 
provides a simple physical explanation for the computational discovery 
that SMS collapse inevitably terminates in the simultaneous 
formation of a SMBH {\it and} a rather substantial 
ambient disk. This disk arises even though the total spin of the progenitor 
star, $J/M^2 = 0.97$, is safely below the Kerr limit.  
The calculation performed here applies to {\it any} marginally unstable
$n = 3$ polytrope uniformly rotating at the break--up speed, independent of
stellar mass or the source of internal pressure.
It illustrates how the black hole and disk parameters can be determined 
for the collapse of other types of stars with different initial density and 
rotation profiles.
\end{abstract}





\section{Introduction}

Recent observations provide increasingly
strong evidence that supermassive black
holes (SMBHs) of mass $\sim 10^6 - 10^{10}M_{\odot}$ exist and that
they are the central engines that power active galactic
nuclei (AGNs) and quasars (Rees 1998, 2001). 
The dynamical formation of SMBHs, as well as the inspiral, collision and
merger of binary SMBHs, are promising sources of long-wavelength 
gravitational waves for the proposed Laser Interferometer Space Antenna (LISA) 
(Thorne 1995; Schutz 2001).  However, the scenario by which SMBHs 
form is still very uncertain (see, e.g., Rees 1984, for an overview).
One promising route is the collapse of a supermassive star (SMS).  
Once they form out of primordial gas, sufficiently massive stars will evolve
in a quasistationary manner via radiative cooling, slowly contracting until
reaching the point of onset of relativistic radial instability. At
this point, such stars undergo catastrophic collapse on a dynamical
timescale,  leading to the formation of a SMBH
(Bisnovatyi-Kogan, Zel'dovich \& Novikov 1966; Zel'dovich \& Novikov
1971; Shapiro \& Teukolsky 1983).

Because most objects formed in nature have some angular momentum,
rotation is likely to play a significant role in the quasistationary
evolution, as well as the final collapse of a SMS. The slow
contraction of even a slowly rotating SMS will likely spin it up to
the mass--shedding limit, because such stars are so centrally
condensed. At the mass--shedding limit, matter on the equator moves in
a Keplerian orbit about the star, supported against gravity by
centrifugal force and not by an outward pressure gradient.
Baumgarte \& Shapiro (1999) recently performed a detailed
numerical analysis of the structure and stability of a rapidly rotating
SMS in equilibrium. Assuming the viscous or magnetic 
braking timescale for angular momentum transfer 
is shorter than the evolution timescale
of a typical SMS (Zel'dovich \& Novikov 1971; Shapiro 2000),
the star will settle into rigid rotation 
and evolve to the mass--shedding limit following cooling and contraction. 
They found that all stars at
the onset of quasi-radial collapse have an  
equatorial radius $R \approx 640GM/c^2$ and a nondimensional
spin parameter $a/M \equiv c J/GM^2 
\approx 0.97$. Here $J$, $M$, $c$, and $G$ 
are spin, gravitational mass, light velocity and gravitational constant. 
(Hereafter we adopt gravitational units and set $c=G=1$).
Because of the large value of $a/M$, 
it was uncertain whether the rotating SMS 
would collapse directly to a black hole and/or form a disk.
Because of the growth of $T/|W| \propto 1/R$ 
during collapse (where $T$ is the rotational kinetic energy and $W$ is the 
gravitational potential energy), it was unclear 
whether the collapse would trigger the growth of bars or other nonaxisymmetric 
instabilities.  
We therefore investigated the collapse 
of a rotating SMS with a $3+1$ post-Newtonian (PN) 
hydrodynamic simulation (Saijo et al. 2002). 
We found that the collapse proceeds nearly homologously and 
axisymmetrically and inferred that a SMBH is likely to be 
formed, with some gas remaining outside the hole. Shibata \& Shapiro (2002)
then performed simulations of the collapse using an axisymmetric code in full
general relativity. We showed conclusively that a black hole forms at the
center and determined that the mass of the hole contains about 
$90\%$ of the total mass of the system and has a
spin parameter $J/M^2 \sim 0.75$. The remaining gas forms a
rotating disk about the nascent hole.

Guided by our numerical simulations,we show here 
that the final black hole mass and spin can be calculated
{\it analytically} (up to quadrature) from the stellar density and angular 
momentum profiles in the progenitor SMS. The calculation 
provides a simple physical explanation for the important finding that 
a SMBH formed from the collapse of a rotating SMS is
always born with a ``ready--made'' disk that can provide fuel for 
accretion to power an energy source.

\section{Basic Assumptions}

Our analysis relies on several explicit assumptions, all of which we expect
to hold to high accuracy in SMSs.  In particular, we assume that the 
SMS is
\begin{enumerate}

\item dominated by thermal radiation pressure;

\item fully convective;

\item uniformly rotating;

\item governed by nearly Newtonian gravitation;

\item described by the Roche model in the outer envelope.

\end{enumerate}

For large masses, the ratio between radiation pressure, $P_r$, and gas
pressure, $P_g$, satisfies
\begin{equation} \label{beta}
\beta \equiv \frac{P_g}{P_r} = 8.49 
	\left( \frac{M}{M_{\odot}} \right)^{-1/2}
\end{equation}
(see, e.g., Shapiro \& Teukolsky 1983, eqs.~(17.2.8) and~(17.3.5));
here the coefficient has been evaluated for a composition of pure
ionized hydrogen.  For SMSs with $M \agt 10^6 M_{\odot}$, we can
therefore neglect the pressure contributions of the plasma in
determining the equilibrium profile, even though the plasma can be
important for determining the stability of the star when it is rotating 
sufficiently slowly (Zel'dovich \&
Novikov 1971; and Shapiro \& Teukolsky 1983).  A simple proof that
SMSs are convective in this limit is given in Loeb \& Rasio (1994).
This result implies that the photon entropy per baryon,
\begin{equation}
s_r = \frac{4}{3} \,\frac{a T^3}{n} 
\end{equation}
is constant throughout the star, and so therefore is $\beta \approx 8
(s_r/k)^{-1}$.  Here $a$ is the radiation density constant, $n$ is the
baryon density, and $k$ is Boltzmann's constant.  As a consequence,
the equation of state of a SMS is that of an $n=3$ polytrope:
\begin{equation} \label{pressure}
P = K \rho^{4/3},  \mbox{~~}
K = \left[ \Big( \frac{k}{\mu m} \Big)^4 \frac{3}{a}
\frac{(1 + \beta)^3}{\beta^4} \right]^{1/3} = {\rm const},
\end{equation}
where $m$ is the atomic mass unit and $\mu$ is the mean molecular
weight (cf.~Clayton 1983, eq.~2-289; note that Clayton adopts a
different definition of $\beta$, which is related to ours by
$\beta_{\rm Clayton} = \beta /(1+\beta)$). In fact, our results
depend only on the assumption that the star is an $n=3$ polytrope,
and not on the origin of the internal pressure.

The third assumption, that the star is uniformly rotating, is probably
the most uncertain of our assumptions.  Nevertheless, it has been
argued that convection and magnetic fields provide an effective
turbulent viscosity which dampens differential rotation and brings the
star into uniform rotation (Bisnovatyi-Kogan, Zel'dovich \& Novikov
1967; Wagoner 1969; Shapiro 2000; for an alternative, 
see New and Shapiro 2001). 
Uniformly rotating configurations were adopted as initial configurations
in Shibata and Shapiro (2002), whose numerical results we are attempting to
corroborate.

We assume that gravitational fields in the progenitor star 
are sufficiently weak so that we
can apply Newtonian gravity in analyzing its envelope.  
SMSs at the onset of collapse have equatorial 
radii $R/M \approx
640$ (Baumgarte \& Shapiro 1999), so that this assumption certainly holds.
Relativistic corrections are crucial for the stability of SMSs, but
can be neglected in the analysis of the equilibrium state.

Finally, the Roche approximation provides a very accurate description
of the envelope of a rotating stellar model with a soft equation of
state, as in the case of an $n=3$ polytrope (for numerical
demonstrations, see, eg, Papaloizou \& Whelan 1973 and 
Baumgarte \& Shapiro 1999).
Since our analysis is based on this approximation, we will briefly
review it together with some of its predictions in the following
section.

We assume that the collapse proceeds
\begin{enumerate}

\item axisymmetrically;

\item adiabatically.

\end{enumerate}

Our previous 3D simulation in PN gravitation (Saijo et al. 2002) allowed
for the growth of nonaxisymmetric perturbations, but none were observed.
Hence the assumption of axisymmetry, which was the basis of the fully
relativistic simulation of Shibata \& Shapiro (2002), appears to be 
justified. Cooling by photon and neutrino emission 
can be ignored on a dynamical timescale (Linke et al 2001), hence the collapse
is essentially adiabatic.  The collapse will be accompanied by a burst of 
gravitational radiation which may be detectable by LISA. However,
the fractional loss of mass--energy is sufficiently small that it can 
be ignored in determining the black hole mass (Shibata \& Shapiro 2002); angular momentum is strictly conserved,
as gravitational waves carry off no angular momentum in axisymmetry.

\section{Review of the Roche Model}

Stars with soft equations of state are extremely centrally condensed:
they have an extended, low density envelope, while the bulk of the
mass is concentrated in the core.  For an $n=3$ polytrope, for
example, the ratio between central density to average density is
$\rho_c/ \bar \rho = 54.2$.  The gravitational force in the envelope
is therefore dominated by the massive core, and it is thus legitimate
to neglect the self-gravity of the envelope.  In the equation of
hydrostatic equilibrium,
\begin{equation} \label{hydro}
\frac{{\bf \nabla} P}{\rho} = - {\bf \nabla} (\Phi + \Phi_c),
\end{equation}
this neglect amounts to approximating the Newtonian potential $\Phi$ by
\begin{equation} \label{potential}
\Phi = - \frac{M}{r}
\end{equation}
In~(\ref{hydro}) we introduce the centrifugal potential $\Phi_c$,
which, for constant angular velocity $\Omega$ about the $z$-axis, can
be written
\begin{equation} \label{centrif}
\Phi_c = - \frac{1}{2} \Omega^2\,(x^2 + y^2) = 
	- \frac{1}{2} \Omega^2 r^2 \sin^2 \theta.
\end{equation}
Integrating eq.~(\ref{hydro}) yields the Bernoulli integral
\begin{equation} \label{bernoulli}
h + \Phi + \Phi_c = H,
\end{equation}
where $H$ is a constant of integration and 
\begin{equation} \label{enthalpy}
h = \int \frac{dP}{\rho} = (n+1)\,\frac{P}{\rho}
\end{equation}
is the enthalpy per unit mass.  Evaluating eq.~(\ref{bernoulli})
at the pole yields
\begin{equation} \label{pole}
H = - \frac{M}{R_p},
\end{equation}
since $h = 0$ on the surface of the star and $\Phi_c = 0$ along the
axis of rotation.  In the following we assume that the polar radius
$R_p$ of a rotating star is always the same as in the nonrotating case. 
This assumption has been shown numerically to be very accurate (e.g.,
Papaloizou \& Whelan 1973). The mass of the star is hardly changed from its
value in spherical equilibrium,
\begin{equation}
M = \frac{4}{\pi^{1/2}} (\xi_1^2 |\theta'({\xi_1})|) K^{3/2},
\end{equation}
where the Lane-Emden factor for an $n=3$ polytrope is given by
$\xi_1^2 |\theta'({\xi_1})|=2.01824$ (Shapiro \& Teukolsky 1983).
A rotating star reaches mass shedding when the equator orbits with the
Kepler frequency.  Using eqs.~(\ref{centrif}) and ~(\ref{bernoulli}), it 
is easy to show
that at this point the ratio between equatorial and polar radius is
\begin{equation}
\left( \frac{R_e}{R_p} \right)_{\rm shedd} = \frac{3}{2}.
\end{equation}
The corresponding maximum orbital velocity is
\begin{equation} \label{shedd}
\Omega_{\rm shedd} = \left( \frac{2}{3} \right)^{3/2}
	\left( \frac{M}{R_p^3} \right)^{1/2}
\end{equation}
(Zel'dovich \& Novikov 1971; Shapiro \& Teukolsky 1983).

Inserting eqs.~(\ref{pressure}),~(\ref{potential}), ~(\ref{centrif}), 
~(\ref{enthalpy}), ~(\ref{pole}) and ~(\ref{shedd}) 
into eq.~(\ref{bernoulli}) yields the density
throughout the extended envelope,
\begin{equation} \label {rho}
\rho = \frac{(\xi_1^2 |\theta'({\xi_1})|)^2}{4 \pi} \frac{M}{R_p^3}
\left(
\frac{R_p}{r}-1 +\frac{4}{27}\frac{r^2}{R_p^2}{\rm sin^2 \theta}
\right)^3.
\end{equation}
The scale of the density is set by the value of the polar radius at the
point of onset of radial collapse, 
\begin{equation} \label {Rp}
R_p/M = 427 
\end{equation}
(Baumgarte \& Shapiro 1999).
The stellar surface is the boundary along which $\rho = 0$, and is thus defined
by the curve $r(\theta)$ given by
\begin{equation}
\frac{4}{27}\frac{r^3}{R_p^3}{\rm sin^2 \,\theta} - \frac{r}{R_p} +1 = 0.
\end{equation}
The solution to this cubic equation is given by
\begin{equation}
\frac {r(\theta)}{R_p}= \frac{3\,{\rm sin \,(\theta/3)}}{{\rm sin \,\theta}}
\end{equation}
(R. Cooper, private communication).

\begin{center}
\begin{minipage}{2.5cm}
\epsfxsize 2.5in
\epsffile{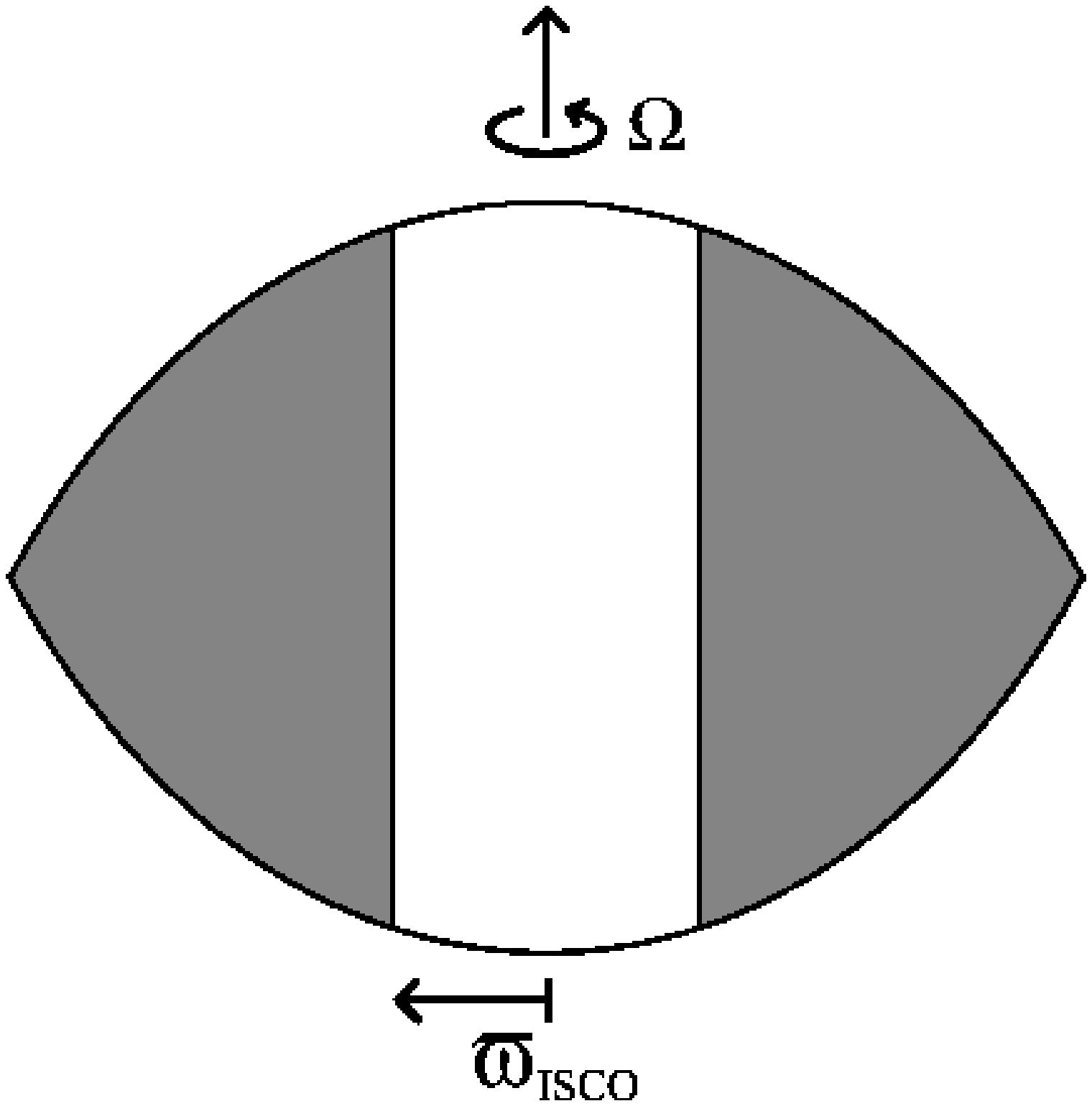}
\end{minipage}
\end{center}
\figcaption[f1.ps]{A SMS rotating uniformly at the mass--shedding limit. 
The shaded region shows
the portion of the Roche envelope $\varpi > \varpi_{\rm ISCO}$ 
where the matter has specific 
angular momentum $j > j_{\rm ISCO}$. Consequently, this gas 
will escape capture by the black hole that forms at the center 
following collapse of the interior.
\label{fig1}}

\section{Calculating the black hole mass and spin}

Consider the implosion of matter from the envelope onto the central
black hole formed from the collapse of the centrally condensed interior 
region.  If the specific angular momentum of the imploding envelope matter 
is below $j_{\rm ISCO}$, the
specific angular momentum of a particle at the innermost stable circular
orbit {\rm (ISCO)} about the hole, the matter will be captured. If the angular momentum of the infalling matter exceeds $j_{\rm ISCO}$,  
the matter will escape capture and continue to orbit outside the hole, 
forming a disk. This capture criterion is well-supported by the numerical
simulation of Shibata \& Shapiro (2002), where  
the capture of fluid in noncircular
orbits with  $j > j_{\rm ISCO}$ was demonstrated to be negligible. 
This criterion suggests a simple iterative scheme
for calculating the final mass and spin of the hole and disk from the initial
stellar density and angular momentum profile. First, 
guess the mass and spin of the hole, $M_h$ and $J_h$.  For 
our initial guess, we shall take a black hole that has consumed
all the mass and angular momentum of the star, so that
$M_h/M = 1 $ and $J_h/M_h^2 = 0.97$. Next, use the initial stellar profile
to ``correct'' this guess by calculating the escaping  mass and angular momentum
of the outermost envelope with specific angular momentum
exceeding $j_{\rm ISCO}$ (see figure 1). We note that
the value $j_{\rm ISCO}$ depends on $M_h$ and $J_h$. We then ``correct'' the 
black hole mass and spin by deducting the 
values of the escaping mass and angular momentum of the envelope material  
from the guessed values of  $M_h$ and $J_h$. We recompute $j_{\rm ISCO}$ for the
``corrected'' black hole mass and spin, and repeat the calculation 
of the escaping envelope mass and angular momentum until convergence is 
achieved. The calculation described above relies on the theorem that for
an axisymmetric dynamical system, the specific angular momentum spectrum,
i.e., the integrated baryon rest-mass of all
fluid elements with specific angular momentum $j$ less than a given value
(e.g., $j_{\rm ISCO}$) is strictly conserved in the absence of viscosity
(Stark \& Piran 1987). Any viscosity, if present, is expected to be unimportant
on dynamical timescales, as required by the theorem.

For a Kerr black hole of mass $M_h$ and spin parameter $a=J_h/M_h$, the
value of $j_{\rm ISCO}$ is given by
\begin{equation} \label{Jisco}
j_{\rm ISCO} =\frac{\sqrt{M_h r_{\rm ms}} (r_{\rm ms}^2 - 2 a \sqrt{M_h 
r_{\rm ms}} + a^2) }{r_{\rm ms}(r_{\rm ms}^2 - 3 M_h r_{\rm ms}
+ 2 a \sqrt{M_h r_{\rm ms}})^{1/2} }
\end{equation}
where $r_{\rm ms}$ is the ISCO given by
\begin{equation}
r_{\rm ms} = M_h \{ 3 + Z_2 - [(3 - Z_1)(3 + Z_1 + 2Z_2)]^{1/2} \}, 
\end{equation}
where
\begin{equation}
Z_1 \equiv 1 + \left 
(1 - \frac{a^2}{M_h^2} 
\right )^{1/3} 
\left [ 
\left (1 + \frac{a}{M_h} \right )^{1/3} +  
\left (1 - \frac{a}{M_h} \right )^{1/3}
\right ],
\end{equation}
and
\begin{equation} \label{Z2}
Z_2 \equiv \left (3 \frac{a^2}{M_h^2} + Z^2_1
\right )^{1/2}.
\end{equation}
(see, e.g., Shapiro \& Teukolsky 1983).
Clearly, the infalling gas corotates with the black hole.  

The mass of the escaping matter in the envelope 
with $j > j_{\rm ISCO}$ is given by
\begin{equation}
\Delta M = \int\!\!\int \, 2\pi \varpi d\varpi \rho
\end{equation}
where the density is given by eq.~(\ref{rho}) and the 
quadrature is performed over all cylindrical shells in the star with
cylindrical radii $\varpi > \varpi_{\rm ISCO} = (j_{\rm ISCO}/\Omega)^{1/2}$
(see fig 1). Here we set $\Omega = \Omega_{\rm shedd}$ 
given by eq.~(\ref{shedd}). 
Defining $\bar{\varpi}=\varpi/R_p$ and $\bar{z}=z/R_p$
gives the nondimensional integral
\begin{eqnarray} 
\Delta M/M &=& \left( \xi_1^2 |\theta'({\xi_1})| \right)^2 \nonumber \\
& & \times \int\!\!\int \, {\bar{\varpi}} d {\bar{\varpi}} d \bar{z} \,
\left [\frac{1}{(\bar{\varpi}^2 + \bar{z}^2)^{1/2}} -1 +
\frac{4}{27}\bar{\varpi}^2
\right ]^3\ . \nonumber \\
& &
\label{dM}
\end{eqnarray}
Similarly, the angular momentum carried off by the escaping 
matter in the envelope is given by
\begin{equation}
\Delta J = \int\!\!\int \, 2\pi \varpi d\varpi \rho \,\Omega \varpi^2 \,
\end{equation}
or
\begin{eqnarray} 
\Delta J/M^2 &=& 
\left( \xi_1^2 |\theta'({\xi_1})| \right)^2
\left( \frac{2}{3} \right )^{3/2}
\left( \frac{R_p}{M} \right )^{1/2} \nonumber \\
& & \times \int\!\!\int \, {\bar{\varpi}}^3 d {\bar{\varpi}} d \bar{z} \,
\left [\frac{1}{(\bar{\varpi}^2 + \bar{z}^2)^{1/2}} -1 +
\frac{4}{27}\bar{\varpi}^2 
\right ]^3\ , \nonumber \\
& & \label{dJ}
\end{eqnarray}
where $R_p/M$ is given by eq.~(\ref{Rp}) and where once again the integral
is performed over all cylindrical shells in the star with
$\varpi > \varpi_{\rm ISCO} = (j_{\rm ISCO}/\Omega)^{1/2}$

The mass and angular momentum of the black hole can then be determined
from eqs.~(\ref{dM}) and ~(\ref{dJ}) according to
\begin{equation} \label{Mh}
M_{h}/M = 1 - \Delta M/M
\end{equation}
and
\begin{equation} \label{Jh}
J_h/M_h^2 = \frac {\left ( J/M^2 - \Delta J/M^2 \right)}
{\left ( 1 - \Delta M/M \right )^2}.
\end{equation}
Once the iteration of 
eqs.~(\ref{Jisco}) -- ~(\ref{Z2}) with
eqs.~(\ref{dM}), ~(\ref{Mh}) and ~(\ref{Jh}) 
converges, the mass
of the ambient disk can be found from
\begin{equation}
M_{\rm disk}/M = \Delta M/M.
\end{equation}

Convergence to better than 1\% is achieved after only four iterations.
We find that $\varpi_{\rm ISCO}/R_p = 0.43$.
We obtain a black hole mass $M_h/M = 0.87$ and black hole spin
$J_h/M_h^2 = 0.71$, and a disk mass $M_{\rm disk}/M = 0.13$.
These values are consistent with those inferred from
the numerical simulation of Shibata \& Shapiro (2002):
$M_h/M \approx 0.9$,
$J_h/M_h^2 \approx 0.75$, and $M_{\rm disk}/M \approx 0.1$.
The differences are well 
within the numerical accuracy of the dynamical integrations in the 
simulation. 

\section{Discussion}

We have considered a SMBH formed by the collapse of a radially
unstable, uniformly rotating SMS of mass $M$ spinning at the mass--shedding 
limit. We calculated that the black hole has a mass $M_h/M \approx 0.87$
and a spin $J_h/M_h^2 \approx 0.71$. Most significant, we found that 
at birth such a  black hole will be  automatically embedded in a 
substantial disk of mass $M_{\rm disk}/M \approx 0.13$. A disk forms even 
though the total angular momentum of the star at the onset of collapse 
is $J/M^2 \approx 0.97$, below the Kerr limit.

We have also calculated the black hole and disk parameters for 
unstable SMSs which, at the onset of collapse, are
spinning more slowly than the angular frequency at break--up 
(i.e. mass--shedding).
We find that the parameters are very insensitive to the ratio $\alpha = 
\Omega/\Omega_{\rm shedd}$, where $\Omega_{\rm shedd}$ is given by
eq.~(\ref{shedd}), over the range $ 0.1 \leq \alpha \leq 1$. The reason is
that for small $\Omega$, a rotating SMS destabilizes at large 
$R_p/M$ ($R_p/M \approx
427/\alpha^2$) with a value of $J/M^2 \approx 0.97$ nearly independent of
$\alpha$. 
The corresponding value of 
$\varpi_{\rm ISCO}/R_p \approx 0.43$ is nearly unchanged, and as a result, 
the final black hole and disk
parameters do not vary significantly as $\alpha$ decreases. (For sufficiently
low spin rates, the effects of thermal gas pressure compete with rotation to
determine the onset of stability and must be included).
Nevertheless, we believe, on evolutionary grounds, that it is most likely 
that a SMS will be spinning
near the break--up value at the onset of collapse (Baumgarte \& Shapiro 1999).

The nascent disk is fairly massive. Such a massive, rotating disk may be
dynamically unstable to nonaxisymmetric instabilities. If the disk
deforms, it could emit long wavelength, quasi-periodic gravitational waves
with a frequency of about $10^{-3} (M/10^6 M_\odot)^{-1}$ Hz, which might 
be detectable by LISA.

At birth, the disk will be hot and thick, 
radiation dominated, and will extend from radius $ \sim 640 M$ at the outer 
edge in the equatorial plane all the way down to the horizon. 
Subsequently, the 
matter in the disk will likely cool by both photon and neutrino emission 
and flatten.
Under the influence of viscosity, the gas will diffuse outwards
in the outermost region, transporting  angular momentum away from the inner 
region.  At the same time, gas in the inner region will accrete 
onto the central black hole, increasing its mass and powering
a source of luminous energy. The lifetime of the disk $t_{\rm disk}$ 
may be estimated crudely by speculating that the accretion eventually will 
settle into a steady state, generating
photons at the Eddington luminosity at $\sim 10\%$ efficiency:
$L = L_{\rm Edd} = 1.3 \times 10^{38} (M_h/M_\odot)$ and 
$\dot{M} = 10L/c^2$. Accordingly, we have 
$t_{\rm disk} = M_{\rm disk}/ \dot{M}$, or
\begin{equation}
t_{\rm disk} = \frac{M_{\rm disk} c^2}{10 L_{\rm Edd}} 
\approx 5 \times 10^6 \, {\rm yrs}.
\end{equation}
The above estimate for $t_{\rm disk}$ is independent of $M$. However, it may
be off by a considerable factor: a lower luminosity will increase the 
disk lifetime, while a lower efficiency (e.g, an efficiency characterizing an 
ADAF disk; Narayan \& Yi 1994) or an enhancement in the total luminosity
via the addition of neutrino cooling, could decrease the lifetime. 
Moreover, the disk at black hole birth may
comprise far more material than the gas which escapes capture
during the final implosion. Prior to undergoing collapse, a rotating
SMS will likely evolve along a mass--shedding sequence as it cools and
contracts, provided the stellar viscosity and/or magnetic braking proves
sufficient to enforce uniform rotation (Baumgarte \& Shapiro 1999). In this
case, the secular cooling and contraction phase 
will be accompanied by mass loss, and the gas which is ejected then
might already form an extensive, ambient disk prior to collapse. Finally we 
note the possibility that such a massive disk around the hole 
may suffer a global ``runaway'' 
instability that may cause a more rapid destruction of the disk than 
implied by steady-state accretion (Abramowicz et al. 1983; Nishida et al. 1996;
Nishida \& Eriguchi 1996). These are 
important issues which are ripe for further analysis.

The  calculation performed here applies to {\it any} marginally unstable 
$n=3$ polytrope rotating at the mass--shedding limit. The results are
independent of stellar mass or the source of internal pressure.
The method should be applicable for estimating
the  black hole and disk parameters for the 
collapse of other types of stars with different
equations of state, initial density and 
rotation profiles. For example, a preliminary calculation indicates that 
for a marginally unstable star uniformly rotating at break--up speed, the
mass fraction that avoids collapse and forms a disk 
falls sharply as the polytropic index of the star drops below
$n=3$. The reason is that such stars are more compact at the onset of collapse,
with smaller values of $R_p/M$ and, hence, less extended envelopes.
This may provide a qualitative explanation for the results 
of previous simulations in full general relativity 
for collapse of rotating neutron stars modeled as
$n=1$ polytropes (Shibata, Baumgarte \& Shapiro, 2000; Shibata, 2000). 
We hope to provide further examples in the future. 

\acknowledgments

We are grateful to the Illinois Undergraduate Research Team 
(H. Agarwal, R. Cooper, B. Hagan and D. Webber)
for assistance with visualization. 
This work was supported in part by NSF Grant PHY-0090310 and 
NASA Grants NAG5-8418 and NAG5-10781 at the
University of Illinois at Urbana-Champaign and 
Monbu-Kagaku-sho Grant Nos. 13740143 and 14047207 
at the University of Tokyo.




\begin{thebibliography}{}
\bibitem{abra83} Abramowicz, M. A., Calvani, M. \& Nobili, L.,
        1983, Nature, 302, 597.
\bibitem{bau99} Baumgarte, T. W. \& Shapiro, S. L., 1999, ApJ, 
526, 941.
\bibitem{bis67} Bisnovatyi-Kogan, G. S., Zel'dovich, Ya. B.,
        \& Novikov, I. D., 1967, Soviet Astron., 11, 419.
\bibitem{clay83} Clayton, D., 1983, {Principles of Stellar Evolution
        and Nucleosynthesis} (University of Chicago Press).
\bibitem{lin01} Linke, F., Font, J. A., Janka, H.-T., M\"uller, E. \&
Papadopoulas, P., 2001, A\&A, 376, 568.
\bibitem{loe94} Loeb, A. \& Rasio, F. A., 1994, ApJ, 432, 52.
\bibitem{new01} New, K. C. B. \& Shapiro, S. L., 2001, ApJ, 548, 439.
\bibitem{nar94} Narayan, R. \& Yi, I., 1994, ApJ, 428, L13.
\bibitem{ni96a} Nishida, S., Lanza, A., Eriguchi, Y. \& Abramowicz, M. A.,
        1996,  MNRAS, 278, L41.
\bibitem{ni96b} Nishida, S. \& Eriguchi, Y., 1996, ApJ, 461, 320.
\bibitem{pap73} Papaloizou, J.~C.~B., \& Whelan, F.~A.~J., 1973,
        MNRAS, 164, 1.
\bibitem{ree84} Rees, M. J., 1984, ARA \& A, 22, 471. 
\bibitem{ree98} Rees, M. J., 1998, 
in {Black holes and relativistic stars}, ed.
R. M. Wald (Chicago University Press),  79.
\bibitem{ree01} Rees, M. J., 2001, 
in {Black holes in Binaries and Galactic Nuclei}, ed.
L. Kaper, E. P. J. van den Heurel, \& P. A. Woudt (New York: 
Springer-Verlap), 351.
\bibitem{sai01} Saijo, M, Baumgarte, T. W., Shapiro, S. L. \&
Shibata, M, 2002, ApJ, 569, 349. 
\bibitem{sch01} Schutz, B. F., 2001, gr-qc/0111095. 
\bibitem{sha00} Shapiro, S. L., 2000, ApJ, 544, 397.
\bibitem{sha83} Shapiro, S. L. \& Teukolsky, S. A., 1983, 
{Black  Holes, White Dwarfs, and Neutron Stars} 
(Wi1ey interscience, New York).
\bibitem{sbs00} Shibata, M., Baumgarte, T. W.
\& Shapiro, S. L., 2000, Phys. Rev. D 61, 044012.
\bibitem{shi00} Shibata, M., 2000, Prog. Theor. Phys, 104, 325.
\bibitem{shib02} Shibata, M. \& Shapiro, S. L., 2002, ApJL, 
in press (astro-ph/0205091).
\bibitem{sta87} Stark, R. F. \& Piran, T., 1987, Comp. Phys. Rep., 5, 221. 
\bibitem{tho95} Thorne, K. S., 1995, in {Proceeding of Snowmass 95
 Summer Study on Particle and Nuclear Astrophysics and Cosmology},
eds. E. W. Kolb and R.  Peccei (World Scientic, Singapore), 398.
\bibitem{wag69} Wagoner, R. V., 1969, ARA\&A, 7, 553.
\bibitem{zel71} Zel'dovich, Ya. B. \& Novikov, I. D., 1971, 
{Relativistic Astrophysics Vol. 1} (University of Chicago Press).
\end{thebibliography}
\end{document}